\documentstyle[aps,twocolumn,epsf]{revtex}
\draft
\begin{document}
\def\bc{\begin{center}}
\def\ec{\end{center}}
\def\btab{\begin{tabular}}
\def\etab{\end{tabular}}
\def\beq{\begin{equation}}
\def\eeq{\end{equation}}
\def\mc{\multicolumn}
\def\ul{\underline}
\def\ol{\overline}
\def\bi{\bibitem}
\def\deg{$^{\circ}$}
\def\C{$^{\circ}$C}
\def\ra{\mbox{$\rightarrow$}}
\def\ef{\mbox{$E_{\rm F}$}}
\def\e3{\mbox{$\epsilon_3$}}
\def\ed{\mbox{$\epsilon_{\rm D}$}}
\def\tc{\mbox{$T_{\rm c}$}}
\def\tmin{\mbox{$T_{\rm min}$}}
\def\rmin{\mbox{$\rho_{\rm min}$}}
\def\rnot{\mbox{$\rho_{\circ}$}}
\def\tnot{\mbox{$T_{\circ}$}}
\def\kb{\mbox{$k_{\rm B}$}}
\def\etal{{\it et al}}
\def\ie{{\it i.e.,}}
\def\eg{{\it e.g.}}
\def\etc{{\it etc.}}
\def\vs{{\it vs}}
\def\wrt{{\it w.r.t.}}
\def\nbs2{NbS$_2$}
\def\2hnbs2{2H-NbS$_2$}
\def\3rnbs2{3R-Nb$_{1+x}$S$_2$}
\def\gaxnbs2{Ga$_x$NbS$_2$}
\def\ganbs{Ga$_{1.33}$Nb$_4$S$_8$}
\def\ganbse{Ga$_{1.33}$Nb$_4$Se$_8$}
%%%%%%%%%%%%%%%%%%%%%%%%%%%%%%%%%%%%%%%%%%%%%%%%%%%%%%%%%%%%%%%%%%%%%%%%
\title{Structural and electronic properties of the Nb$_4$-cluster compound\\ 
Ga$_{1.33}$Nb$_4$X$_8$ (X = S, Se)}

\author{Asad Niazi\protect$^{1}$ and A. K. Rastogi\protect$^{2}$}
\address{\protect$^{1}$ Tata Institute of Fundamental Research, Homi Bhabha Road, 
Mumbai - 400005, India \\
\protect$^{2}$ School of Physical Sciences, Jawaharlal Nehru University,
New Delhi - 110067, India}

\maketitle

\footnotetext[1]{Present Address: Dept. of Physics, Faculty of Natural
Sciences, Jamia Millia Islamia, New Delhi 110025, India. Email: asad@jamia.net} 
 
\footnotetext[2]{Email: akr0700@mail.jnu.ac.in}

\begin{abstract}
We report resistivity, thermopower and magnetic susceptibility measurements
on the Nb$_4$-cluster compounds Ga$_{1.33}$Nb$_4$X$_{8}$ (X = S, Se),
derived from vacancy ordered spinels A$_x$T$_4$X$_8$. The cubic selenide phase 
is insulating and its resistivity crosses over from $\ln{\rho} \sim T^{-1}$
to $\sim T^{-1/2}$ on cooling below 150~K, indicating variable range
hopping (VRH) conduction at low temperatures. This is similar to that
previously reported for V$_4$ and Mo$_4$ cluster compounds. The
rhombohedrally distorted sulfide is metallic and shows a minimum in
resistance at $\sim $ 56~K below which the resistivity varies as $\rho(T)
\sim T^{-1/2}$. The thermopower of the selenide becomes temperature
independent below the crossover temperature. These Nb$_4$ compounds
exhibit enhanced Pauli-paramagnetic magnetic susceptibility ($\chi$)
irrespective of their transport properties and both undergo a similar transition
in $\chi(T) \sim 30$~K. We discuss these properties in the model of hopping
conduction under long-range Coulomb repulsion effects and derive
consistency between some of the transport and magnetic parameters.
\end{abstract}

%70. Condensed matter: Electronic structure, electrical, magnetic, and optical
%properties.\\
%72.20.-i -- Conductivity phenomena in semiconductors and insulators.\\
%72.20.Ee -- Conductivity phenomena in semiconductors and insulators: Mobility edges;
%hopping transport.\\
%72.80.Ga -- Conductivity of specific materials: Transition metal compounds.\\
%75.20.-g -- Diamagnetism, paramagnetism, and superparamagnetism.

%%%%%%%%%%%%%%%%%%%%%%%%%%%%%%%%%%%%%%%%%%%%%%%%%%%%%%%%%%%%%%%%%%%%%%%%%%%%

\section{Introduction}
Cation rich compounds of the early transition metals, especially the group
4$d$ and 5$d$ chalcogenides, show large clustering of the metal sublattice
with groups of metallic-bonded atoms separated by large intercluster
distances. These  structural features remarkably increase electronic
correlations and lead to a variety of phase transitions at low temperatures.
Of special interest are compositions such as AT$_6$X$_8$ and
AT$_4$X$_8$, where high-symmetry cubic and hexagonal-rhombohedral
structures are obtained and the cluster disorder is expected to be small. The
metallic phases of Mo$_6$ and Rh$_4$ have been found to be
superconducting, the \tc\ increasing with intercluster distances. The 
superconducting parameters suggest a highly localised wavefunction on the
atomic clusters \cite{vandenberg,fischer}. For still larger intercluster
distances such as in V$_4$, Nb$_4$ and Mo$_4$ compounds, the transport
properties are found to be dominated by Coulomb and exchange interaction
effects among cluster electrons, as even a small disorder in the lattice
potential gives a localised electron glass state. They also become magnetic
below about 30~K \cite{sahoo1,sahoo2,asad1,rastogi1}. The compounds of
the present study belong to the same family and are expected to show
smaller correlation effects compared to V$_4$ and Mo$_4$.

All the compounds AT$_4$X$_8$ (A = Ga, Al; T = V, Nb, Ta, Cr and Mo;  X
= S, Se, Te, space group $F{\ol 4}3m$) are insulating and are derived  from
spinels AB$_2$O$_4$ (space group $Fd3m$) with vacancy ordering at the
A-sites  and metal-clustering (T$_4$) of B-site cations. The dominating
effect of metal  clustering on the electronic properties is seen in the
remarkably similar transport and magnetic properties among the compounds
of V$_4$ and Mo$_4$ belonging to the 3$d$ and 4$d$ series respectively
\cite{sahoo1,sahoo2,asad1,rastogi1}. The magnetic properties of Nb$_4$,
Ta$_4$ and mixed Nb$_4$-Mo$_4$ compounds have been reported in a
previous study \cite{rastogi2}. The large Pauli-paramagnetism measured in
Nb$_4$ suggested a partially occupied, although localised electronic band
of states at the Fermi-energy (\ef). Recent band structure calculations on
V$_4$ and Mo$_4$ compounds also gave a partially empty and very narrow
$d$-band at \ef\ which is in accord with their magnetic properties and the
observed hopping conduction of carriers \cite{ddsarma}. The mechanism of
electron localisation, however, remains unexplained.

In this paper we present the structural, transport and magnetic properties of
\ganbs\ and \ganbse. While both these phases were previously reported to
be cubic at room-temperature \cite{rastogi2}, our sulfide compound reported
here is found to have rhombohedral symmetry and shows metallic conduction
in contrast to insulating  behaviour of the cubic phases. However, its lattice
constants, the observed superstructural phases and the remarkable similarity
in its magnetic properties with the cubic phase suggest important similarities
in the short range atomic arrangements in the metallic and the insulating
phases.

\section{Synthesis and structure}
A series of \gaxnbs2 ($x = 0.1, 0.25, 0.33$) was prepared from the elements
in  sealed quartz tubeing 5\% excess sulfur at 850~\C\ (low temperature (LT)
phase) and also above 1100~\C\ (high temperature (HT) phase). For the
selenide the starting composition used for reaction of the elements was
Ga$_{.33}$NbSe$_2$ (+ 5\% excess Se) and final sintering of pellets was
done at 950~\C. Single crystal flakes could be obtained for \2hnbs2 and for
3R-Ga$_{.1}$\nbs2. The details are given elsewhere \cite{asad2}.

\subsection{3R-\gaxnbs2 ($x = 0 - 0.33$)}
Here we briefly review the properties of 3R-\gaxnbs2. Structurally, all these
phases except for $x = 0.33$ (HT), are derived from an intercalated \3rnbs2
($x \le 0.1$) host. The intercalated atoms occupy octahedral sites in the van
der Waals gap giving a rhombohedral unit cell with $R3m$ symmetry. The
transport properties of pressed sintered pellets and the flakes showed metallic
behaviour and a resistance minimum between 20--60~K. The resistance
minimum was related to the combined effects of increasing disorder and
clustering interactions among in-plane Nb atoms \cite{asad2}.

\begin{figure}
\centerline{\epsfxsize=7.5cm{\epsffile{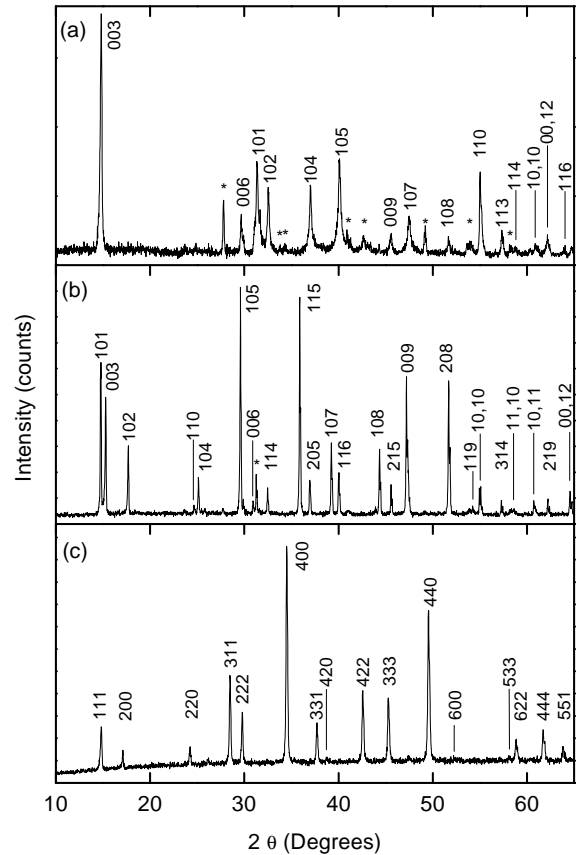}}}
\caption{Powder x-ray diffraction patterns of (a) 3R-Ga$_{0.33}$NbS$_2$
(LT), (b) 3R-Ga$_{0.33}$NbS$_2$ (HT), and (c) {\it f.c.c.} \ganbse. (a) The
superlattice lines are marked by ($\star$) in the LT sulfide. (b) The HT sulfide
has a $2 \times 2$ hexagonal supercell, and is referred to as \ganbs\ in the text.}
\label{fig1}
\end{figure}

For 33\% Ga, the structure of the 1100~\C\ quenched HT phase was found
to be markedly different from the LT phase prepared below 850~\C. This is
evident on comparing their XRD patterns in Fig.~\ref{fig1}(a) and (b). The
structural parameters of the LT phase, $a_H$ = 3.34~\AA, $c_H$ =
7.90~\AA, are similar to the other Ga-intercalated NbS$_2$ compounds.
However, there is a large increase in the number and intensity of the
superstructural lines (marked by $\star$ in Fig.~\ref{fig1}(a)) indicating
increased clustering of Nb atoms in the $a$-$b$ plane. The HT phase has a
qualitatively different structure. All the lines in the pattern can be indexed on
a $2 \times 2$ hexagonal supercell, but with a substantially increased
corresponding $a$-parameter of $2 \times 3.6$~\AA. There is also a large
reduction in its $c$-parameter to a value of 17.31~\AA, making $c/a =
1.603$, as compared to $\sim 1.80$ for the rest of the compounds. The HT
phase remains metallic in spite of these structural changes. However, as we
later show, its magnetic properties are changed and become remarkably
similar to the insulating phases of Nb$_4$-compounds, suggesting similar
clustering in our HT phase.

The rhombohedral cell parameters ($a_R$, $\alpha$) of our HT phase and
that of the cubic GaNb$_4$S$_8$ ($a_C = 10.02$~\AA) reported earlier
\cite{rastogi2} are very similar, being (7.11~\AA, 60.81\deg) and (7.085~\AA,
60\deg) respectively. It is therefore worthwhile to compare other important
features of rhombohedral layered CdCl$_2$ and cubic spinel structures from
which our Ga$_{.33}$\nbs2 and GaNb$_4$S$_8$ are respectively derived.
In both cases the anions form a cubic packing. In CdCl$_2$, the transition
metal atoms occupy all the intralayer-octahadral sites, which alternate with
the intercalate atoms in the van der Waals gap. In spinels on the other hand,
1/4 of the transition metal atoms are transferred to the octahedral sites in the
van der Waals gap so as to form (a) a $2 \times 2$ vacancy ordered
supercell and (b) a tetrahedral arrangement of neighbouring transition metal
atoms. The clustering interactions in our compounds give an {\it f.c.c.} lattice
of well-separated T$_4$ clusters. We believe that in Ga$_{.33}$\nbs2 this
arrangement is obtained by high temperature synthesis. There may be
differences in the site (tetrahedral/octahedral) occupancy of Ga atoms in
these phases which can be resolved by EXAFS studies. We shall therefore,
in the subsequent discussion represent 3R-Ga$_{.33}$\nbs2 as \ganbs. 
    
\subsection{\ganbs\ and \ganbse}
As mentioned above, high temperature synthesis of sulfides for 33\% Ga
gave a rhombohedrally distorted phase whose structure is quite similar to the
previously reported cubic Nb$_4$-cluster compound. In contrast, the
selenide compound was easily obtained with a cubic structure as can be
seen from its XRD pattern reported in Fig.~\ref{fig1}(c). The presence of
(200), (420) and (600) lines indexed on an {\it f.c.c.} lattice ($a_C =
10.41$~\AA) indicates the $F{\overline 4}3m$ space group. A detailed single
crystal analysis of the cubic phases \cite{rastogi2} had shown the existence
of strong tetrahedral Nb$_4$-clustering. In common with the V$_4$ and
Mo$_4$-cluster compounds, the intracluster distances were found to be
3.026~\AA, similar to Nb metal, but large intercluster distances of 4.320~\AA\
rendered these compounds insulating.

{\bf A note on polymorphism:} The crystal chemistry of ternary chalcogenides
of the early transition elements is dominated by CdI$_2$-NiAs-type covalent
structures. There are only a few cubic thiospinels of Ti, V and Zr and they
invariably contain Cu on tetrahedral sites. The existence of cubic structures
for A$_x$TX$_2$ (A = Ga, Al, Cu; T = early transition element) seems to be
intimately related to the extraordinary clustering of transition elements on the
T-site. It should be noted that the more ionic compositions of late transition
elements of the 3$d$ series (Cr--Ni) can exist as cubic thiospinels as well
as subtractive NiAs-type hexagonal modifications whether prepared
synthetically or obtained as naturally occurring minerals. Polymorphic
transformation to the latter variety is also obtained at high pressure and
temperature when the A-site has unfilled orbitals or has equal preference for
octahedral sites of the structure \cite{hulliger}. In our Ga$_x$Nb$_4$S$_8$
compounds too we observe similar structural modifications of phases upon
using different methods for their synthesis. Similarly, in a recent study of
Cu-substituted GaV$_4$S$_8$, cubic, hexagonal and orthorhombic
structures were obtained using different methods of synthesis -- from the
elements and by reduction of oxides at different temperatures. The hexagonal
phases were found to be metallic, while cubic phases showed insulating
behaviour and interestingly the orthorhombic ones underwent metal to
insulator transition below 180~K \cite{majid}. 

\section{Transport Properties}
\subsection{Metallic \ganbs}
\begin{figure}
\centerline{\epsfxsize=7.5cm{\epsffile{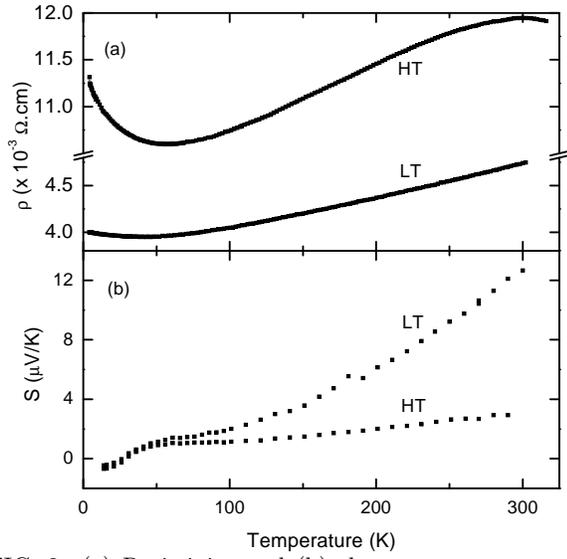}}}
\caption{(a) Resistivity and (b) thermopower versus temperature in the LT and
HT phase of metallic \ganbs. (a) The HT phase shows a maximum in
$\rho(T)$ around room temperature and a prominent minimum $\sim 56$~K. In
comparison, the LT phase shows a monotonic increase in $\rho(T)$ above a
shallow minimum $\sim 40$~K. (b) The $S(T)$ above 100~K also shows
contrasting behaviour in the two phases.}
\label{fig2}
\end{figure}

The structural and magnetic properties of cubic (insulating) phases of
Nb$_4$ compounds have been reported previously \cite{rastogi2}. We now
compare the conductivity and thermopower measurements on pressed
sintered pellets of HT and LT \ganbs\ in Fig.~\ref{fig2}. Both these
compounds show metallic behaviour on cooling and a resistance minimum
respectively at about 40~K and 56~K. A distinct anomaly is also noticed in
the thermopower below \tmin\ where a rapid drop from positive to negative
value occurs. There is however a marked difference in the overall behaviour
related to the increased atomic clustering in the HT phase. 

The HT phase has larger resistivity, and it saturates and passes through a
broad maximum above room temperature. On the other hand, the resistivity
variation of the LT phase is quite similar to and properly scales with the
variations found in the lower Ga-containing phases including \3rnbs2,
observed on plotting $\rho/\rmin$($T/\tmin$) between $.2<T/\tmin<2$
\cite{asad2}. This scaling is obtained for resistivity values differing by an
order of magnitude. The changed $\rho(T)$ behaviour compared to the LT
phase is certainly related to structural changes and significantly increased
electron localisation upon metal-atom clustering in the HT phase.  

The increase in carrier localisation can also be seen in the thermopower of
the HT phase which remains independent of temperature above \tmin,
whereas in the LT phase there is rapid increase on heating. It is here
significant to note that upon cooling below \tmin\ all the phases including
Nb$_{1+x}$S$_2$ showed a similar changeover of $S$. The overall
thermopower behaviour of these compounds is quite complicated and a
satisfactory explanation of the different behaviour above \tmin\ cannot be
provided. A temperature independent thermopower along with resistance
maximum at high temperature is expected for polaronic transport. However,
the band parameters and significant electron-electron interaction effects
causing resistance minimum at low temperatures in these compounds are
inconsistent with the simple model of non-degenerate gas of small polarons,
as suggested for many doped transition metal oxides \cite{mott}. We, on the
other hand, believe that in these cluster compounds the electronic properties
are primarily related to the lattice disorder and strong correlation effects
among the carriers.

\subsection{Insulating \ganbse}
\begin{figure}
\centerline{\epsfxsize=7.5cm{\epsffile{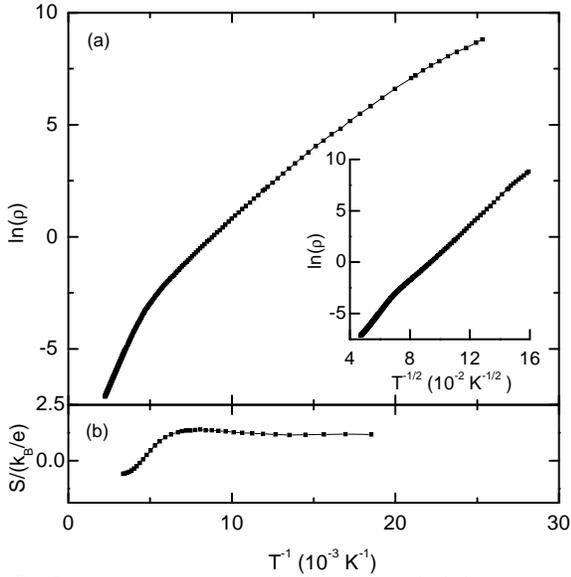}}}
\caption{Temperature variation of resistivity and thermopower in insulating
\ganbse. (a) $\ln{\rho}$ versus $(1/T)$, shows non-Arrhenius behaviour
over the whole $T$ range; (b) $S/(k_{B}/e)$ versus $(1/T)$, is temperature
independent below 150~K. In the inset, $\ln{\rho}$ versus $(1/T)^{1/2}$ is
linear below 150~K.}
\label{fig3}
\end{figure}

As already mentioned, in common with the other tetrahedral cluster
compounds, the cubic phase of our \ganbse\ is insulating. In Fig.~\ref{fig3}(a)
we show its resistivity behaviour between 40--500~K on a $\ln{\rho}(1/T)$
plot. Figure~\ref{fig3}(b) shows the thermopower behaviour between
50--300~K on a $S(1/T)$ plot. The thermopower changes sign from a
negative value at room temperature to positive below 200~K. A general
expression of diffusive thermopower, irrespective of the nature of conduction
(band or hopping conduction), can be written as 
\beq
S = {-k_{\rm B}\over{|e|}}({<\epsilon>-\mu\over{k_{\rm B}T}})
\eeq
where $<\epsilon>$ is the average energy and $\mu$ is the chemical
potential. For an extrinsic semiconductor, ($<\epsilon>-\mu$) would
represent the energy of carrier generation; and for band conduction similar
slopes are obtained for $\ln{\rho}(1/T)$ and $S(1/T)$ plots. The results
presented in Fig.~\ref{fig3} clearly show that the activation energy for
electrical conduction is quite different from that observed in thermopower.
The large difference in slope at high temperature cannot be accounted for by
polaronic hopping. Moreover, there is a strong downward curvature in the
resistivity plot while the thermopower becomes constant below 150~K. Thus,
the activation energy reduces from $\sim 0.14$~eV at high temperature to
just $\sim 0.026$~eV at 40~K. These results are qualitatively similar to those
obtained earlier for the V$_4$ and Mo$_4$-cluster compounds
\cite{sahoo1,asad1}.

We shall later analyse some of the results in the model of impurity band
conduction in a semiconductor, where variable range hopping (VRH)
observed at low temperatures has been extensively studied
\cite{mott,shklovskii}. Moreover, as shown in the inset, the resistivity follows
$\rho = \rnot\exp{(\tnot/T)}^{1/2}$ below 150~K, while as mentioned above,
the thermopower becomes nearly temperature independent. A value of
$\tnot=1.8\times 10^{4}$~K is obtained from the plot. The $T^{-1/2}$
dependence has been previously reported for the cluster compound
GaMo$_4$Se$_4$Te$_4$ where a crossover from $\ln{\rho} \sim T^{-1/4}$
to $\ln{\rho} \sim T^{-1/2}$ was found below a similar temperature of 175~K.
This behaviour was attributed to Coulomb repulsion effects on the VRH of
carriers at low temperature \cite{asad1}. Taking into account the Coulomb
gap effects near the \ef\ of hopping carriers, Burns and Chaikin have earlier
obtained an expresion for $S$ that tends to a constant value at low
temperatures \cite{burns}.

\section{Magnetic Properties}
The magnetic properties of different cluster compounds of V$_4$, Nb$_4$,
Ta$_4$ and Mo$_4$ have previously been reviewed \cite{rastogi3}. The
ferromagnetism in V$_4$ and Mo$_4$ compounds, which also show VRH
conduction, had been discussed in the model of Anderson localised states
by including on-site Coulomb repulsion and exchange interaction among
singly occupied sites \cite{lamba}. In the following we present paramagnetic
properties of \ganbs\ and \ganbse. 

\begin{figure}
\centerline{\epsfxsize=7.5cm{\epsffile{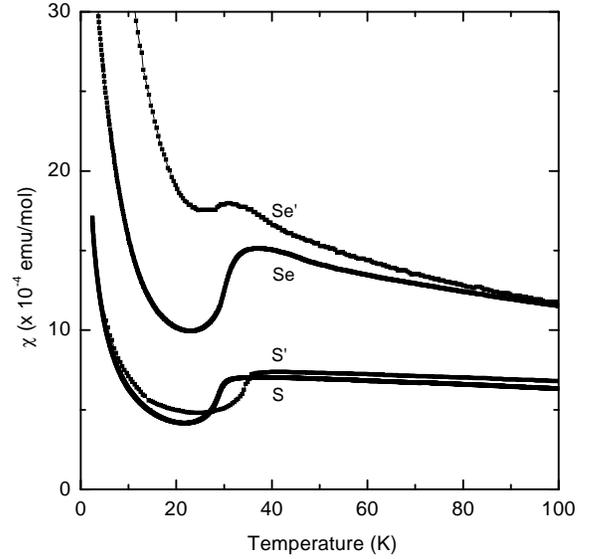}}}
\caption{Magnetic susceptibility of metallic \ganbs\ (S) and insulating
\ganbse\ (Se) plotted along with that of insulating GaNb$_4$S$_8$ (S') and
GaNb$_4$Se$_8$ (Se') phases (ref.~\protect\cite{rastogi2}). The metallic as
well as insulating phases show similar behaviour at and below the transition
around 30~K.}
\label{fig4}
\end{figure}

In Fig.~\ref{fig4}(a), we present the magnetic susceptibility ($\chi(T)$) results
of Nb$_4$-compounds. Qualitatively similar behaviour on cooling, including
a sharp reduction in $\chi$ around 30~K followed by a Curie tail can be
seen in all of them including the metallic sulfide phase of the present study.
The $\chi$ of the selenide is about twice the value of the sulfide indicating
that the partially filled band has larger DOS(\ef), principally due to larger
intercluster distances in selenides. The susceptibility in these compounds
should be compared to much weaker paramagnetism as well as the absence
of low temperature transition in Nb$_4$Se$_4$I$_4$ \cite{rastogi2}. In this
compound similar tetrahedral Nb$_4$ clusters are formed, but now eight
valence electrons per cluster would  correspond to a full $d$-sub band. 

It is important to note that in Nb$_4$-cluster compounds, $\chi$ shows a fall
below the transition whereas in V$_4$ and Mo$_4$ compounds the $\chi$
increases sharply below the transition around 45--55~K and they become
ferromagnetic below 10--25~K. However, there is no qualitative difference
in their transport properties as all of them show VRH conduction. The
remarkable independence of magnetic properties from their transport
behaviour is also seen in Fig.~\ref{fig4} where the metallic and insulating
phases of sulfides are seen to have the same magnitude of $\chi$. We also
note that in the metallic phase, except for a minimum at a higher temperature
of 56~K, no visible anomaly is noticed in the resistance corresponding to the
sharp decrease in $\chi$ below 30~K. In earlier studies, V$_4$ and Mo$_4$
compounds also showed only a weak cusp in ac and dc-resistance in their
$\ln{\rho}$ vs $1/T$ plots around 45--55~K corresponding to the transition
in $\chi$ \cite{sahoo1,sahoo2}. Similar studies for the transport properties of
our insulating \ganbse\ could be not be made since its dc-resistance
became very large below 40~K. 

\begin{figure}
\centerline{\epsfxsize=7.5cm{\epsffile{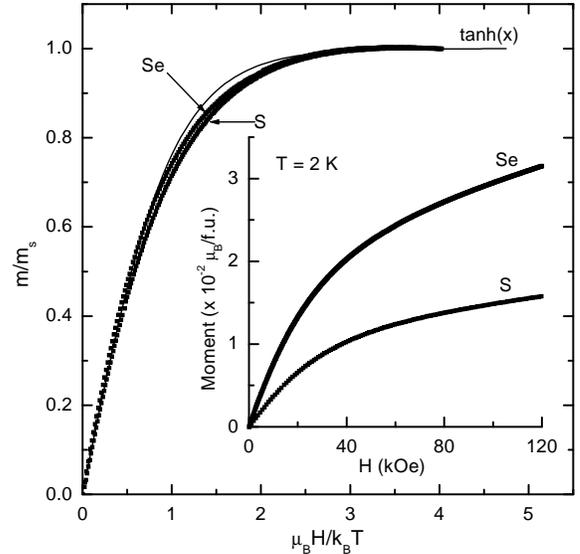}}}
\caption{The local moment contribution of metallic \ganbs\ and insulating
\ganbse. It closely follows the Langevin function (spin 1/2) which is shown
by the solid curve. The inset shows the magnetisation at 2~K uptil a field of
120~kOe from which the above data is extracted.}
\label{fig5}
\end{figure}

The $\chi(T)$ in all the Nb$_4$-compounds showed a Curie tail at low
temperatures. In the inset of Fig.~\ref{fig5}, we have shown $M(H)$ plots of
metallic \ganbs\ and insulating \ganbse\ at 2~K upto 120~kOe field. Both
show a characteristic curvature and a linear slope, indicative of localised
moments ($x$/mole) and high field Pauli-paramagnetic contribution
($\chi_{HF}$) respectively. In the main figure we have replotted the local
moment contribution and compared it to a Langevin function for spin 1/2
moments for both the compounds. This simple analysis gives the following
parameters : $x = 1.02 \times 10^{-2}$ and $1.91 \times 10^{-2}$/mole and
$\chi_{HF} = 1.74 \times 10^{-4}$ and  $3.86 \times 10^{-4}$~emu/mole
respectively for \ganbs\ and \ganbse. 

\begin{figure}
\centerline{\epsfxsize=7.5cm{\epsffile{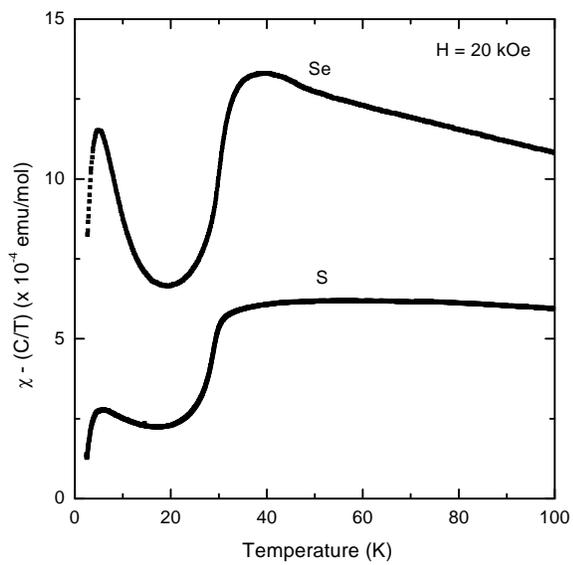}}}
\caption{Temperature variation of $\chi$ in metallic \ganbs\ and insulating
\ganbse, obtained by subtracting the Curie contribution from the data of
Fig.~\protect\ref{fig4}. The further downward trend below 6~K in our
post-subtraction data could be an artifact of $\chi$ measurement at a high
field of 20~kOe.}
\label{fig6}
\end{figure}

The values of $x$ obtained above from the saturated spin-1/2 moments give
Curie constants equal to $3.83 \times 10^{-3}$, $7.16 \times 10^{-3}$
respectively for the sulfide and selenide. These values are used to subtract
the Curie contribution from the $\chi(T)$ results of Fig.~\ref{fig4}, and the
susceptibility is plotted in Fig.~\ref{fig6}. It is uncertain whether a sharp
change in $\chi(T)$ below 6~K is an artifact of our subtraction procedure or
an intrinsic property of these compounds. Apart from this, the results obtained
are very similar to those reported earlier for the Nb$_4$ compounds. The
value of $x \sim 0.01$ in these phases is sufficiently large not to be attributed
to extrinsic magnetic impurities. Rather, these localised moments seem to be
related to the local changes in the Nb$_4$-lattice arising from
non-stoichiometry. This conclusion is in accord with the results on mixed
clusters of Nb-Mo, where the Curie constant reduces from the spin-1/2 per
cluster in Mo$_4$ to about 1--2\% of the value in Nb$_4$-compounds. It is
significant to note that while V$_4$ and Mo$_4$ compounds showed a large
increase in $\chi(T)$ at 45--55K, Nb$_4$ compounds showed a drop
around 30~K. The reason for this difference is not clear at the moment. 

\section{Discussion}
The most significant result of the present study is the synthesis of a
rhombohedral \ganbs\ phase which is found to be metallic, while the cubic
phases of GaNb$_4$S$_8$, GaNb$_4$Se$_8$ and \ganbse\ are insulating.
There is however remarkable similarity in their magnetic properties. The
change to metallic behavior cannot be simply related to carrier doping by
impurity atoms as in a wide band semiconductor. The occurance of metallic
conduction in a rhombohedrally distorted Nb$_4$ compound clearly
suggests important changes in band-dispersion due to cluster distortion and
consequently in the condition for Anderson-localisation of the
wave-function. The disorder effects, however, remain significant as seen
from the $R(T)$ and S$(T)$ anomalies in the metallic phase. The observed
$\sigma(T) \sim T^{1/2}$  below \tmin\ has been related to the Coulomb gap
corrections to the DOS in disordered solids as suggested by Altshuler and
Aronov \cite{altshuler}. In the cubic phases of Nb$_4$, V$_4$ and Mo$_4$,
the transport and magnetic properties are also insensitive to the changes in
concentration or substitution of Ga or Al by Cu, Zn, \etc. This insensitivity to
carrier doping indicates that in cluster compounds the Fermi level is pinned
to a band of states. The states are localised due to disorder and correlation
effects and conduction is by phonon assisted hopping of carriers. A survey
of the high temperature activation energy ($\e3/\kb$) for different sulfides,
selenides and tellurides of cluster-compounds gives values between
700--1800~K: The highest for GaV$_4$S$_8$ and lowest for
GaMo$_4$Se$_4$Te$_4$ \cite{sahoo1,sahoo2,asad1}. These observations
indicate that the localisation of the wavefunction of cluster electrons has
similar origin for different transition metal atom clusters. All the insulating
compounds show strong downward curvature in the $\ln{\rho}(1/T)$ plots,
which is characteristic of Mott's VRH conduction. In some cases Coulomb
repulsion effects on the hopping carriers give a changeover to  $\rho =
\rnot\exp{(\tnot/T)}^{1/2}$ behavior at low temperature, as we have reported
above for \ganbse\ below 200~K. 

For cubic structures, recent electronic energy calculations based on Density
Functional Theory within a local density approximation clearly show that the
ideal thiospinel structures of these compounds are metallic and that there is
substantial effect of atomic clustering on the band energies near the Fermi
level \cite{ddsarma}. The dominant states near the \ef\ are from the
$d$-orbitals of transition elements, and upon their clustering there appears a
gap $\sim 1.4$~eV above the occupied band. Surprisingly, the Fermi level
now remains in a much narrower lower $d$-band leaving hole-like states
just below the band-gap. The wavefunctions of these empty states are
expected to be drastically modified by subtle electronic correlations and also
by small atomic disorder in actual compounds, leading to Mott's
metal-insulator (M-I) transition. The DOS(\ef) for different compounds of
V$_4$ and Mo$_4$ is calculated to be substantially high and varies
between 10--20 states/eV/F.U. \cite{ddsarma}. Qualitatively similar
conclusions about high DOS(\ef) were drawn from the detailed analysis of
low temperature magnetic, specific heat and transport properties
\cite{sahoo1,asad1}.

We observe a similarity of transport properties, especially the crossover to
$\ln{\rho} \sim T^{-1/2}$ behavior below 200~K for \ganbse, with
GaMo$_4$Se$_4$Te$_4$ reported earlier \cite{asad1}. This shows the
importance of correlation effects due to Coulomb repulsion among hopping
electrons as suggested for the impurity doped, moderately compensated
semiconductors \cite{shklovskii}. A detailed analysis of the transport and
magnetic parameters using a similar model will be reported in future. In the
following analysis we try to check the consistency of parameters in the
above model. For \ganbse, $\e3/\kb = 1700$~K, Shklovskii parameter $\tnot
= 1.85 \times 10^4$~K and $\chi_{300}= 12\times 10^{-4}$~emu/mol. 

We should note that for our cluster-compounds the suitability of the model for
hopping conduction among widely separated impurity centers in a lightly
doped-compensated Si or Ge semiconductors rests on the important
observation by Hill that, for a wide variety of solids showing VRH, the
normalised site density, \ie\ $N^{1/3}\alpha^{-1}$ ( N = density of sites and
$\alpha^{-1}$ = localisation length of carriers) falls within a narrow range of
0.1--0.14 \cite{hill}. In our cluster compounds this condition seems to be
satisfied even under a high density of carriers. 
 
For impurity bands a random electrostatic potential originates from
fixed complexes of ionised donors and acceptor impurities. This gives
a dispersion of band energy of the impurity states around a maximum
which is located near the energy of non-ionised donors. The DOS falls
to zero within the characteristic ionisation energy of donors. The
chemical potential (\ef), calculated for lightly compensated
impurities, is found at an energy $\ef=0.61~\ed$ away from the maximum;
where $\ed = e^2/\kappa r_d$, $\kappa$ = dielectric constant and $r_d$
the average separation of donors, \cite{shklovskii}. For nearest
neighbour hopping at high temperatures the conductivity is due to
ionisation of donors located near the maximum of DOS. Thus a constant
activation energy is obtained, given by 
\beq
\e3 = \ef =.61~\ed.
\label{eqn1}
\eeq
In case of our cluster compounds, similar compensation effects can
originate from the Ga atoms occupying tetrahedral sites and may also
contribute the needed random potential for the dispersion of band
energies. In this context, it is important to recall that here the
tetrahedral sites of spinels are only partially occupied and occupancy
disorder may be significant enough to cause localisation of cluster
electrons.

At low temperatures, hopping is among impurities having their energies
near the Fermi level, and Mott's VRH is obtained for a flat DOS.
However, the electronic correlations will modify DOS and hence hopping
conduction. A qualitative estimate of $g(\ef)$ can be made in case of
low compensation. The DOS falls off to zero within \ed\ and the empty
states in the band should correspond to the density of compensating
impurities, thus giving $N_a \sim \frac{1}{2}g(\ef)\ed$, i.e. $g(\ef)
\sim 2N_a/\ed$.

Various studies on the effect of long range Coulomb interaction between
hopping electrons have shown that the single particle DOS vanishes at
\ef\ with a soft gap $\Delta$. This so called Coulomb gap has important
effects on the conduction below a temperature $\Delta/\kb$, where for
hopping in 3-dim, Mott's $\ln{\rho}\sim T^{-1/4}$ is replaced by $\sim
(\tnot/T)^{-1/2}$ dependence. The following expressions have been
derived in case of doped-compensated semiconductors by Efros and
Shklovskii \cite{shklovskii}:
\beq
\kb\tnot = 2.8 e^2/\kappa\alpha^{-1},
\label{eqn2}
\eeq
\beq
\Delta^{2/3}=\ed[r_d g(\ef)^{1/3}].
\label{eqn3}
\eeq
Let us now apply these relations in case of \ganbse. The observed activation
of 1700~K around room temperature and intercluster distance of 4.3\AA\ for
Nb atoms would require a dielectric  constant $\kappa=14$ from
relation~(\ref{eqn1}). This value of $\kappa$ should be compared with about
25 expected for Se$^{2-}$ ions. In another way the parameter
$e/\sqrt{\kappa}$ in (\ref{eqn1}) can be thought of as effective charge of the
hopping carriers. Using $N_a$ as the Ga concentration, we obtain $g(\ef) =
8.5$~states/eV/F.U. . This value should be compared with the band structure
calculations of 10--20~states/eV/F.U. for various cluster compounds
\cite{ddsarma}. The expected magnetic susceptibility is therefore $\sim 1.4
\times 10^{-4}$~emu/mol. The observed value of $6 \times 10^{-4}$ at
300~K for our Nb$_4$ compound shows a significant exchange enhanced
paramagnetism. The compounds of V$_4$ and Mo$_4$ become magnetic
due to much larger exchange correlation among the clustered electrons
\cite{rastogi3}. Using relation~(\ref{eqn2}), the Efros-Shklovskii parameter
\tnot\ can be used to calculate the localisation length $\alpha^{-1}$ which
equals about 1.8~\AA. However, relation~(\ref{eqn3}) for the Coulomb gap
does not seem to give the observed crossover temperature of about 200~K,
as we obtain a value larger by an order of magnitude. This in our opinion is
quite significant and certainly indicates a different nature of correlated
hopping in these compounds. It should be remembered that the above model
of hopping conduction in doped-compensated semiconductor considers
fixed impurity centres, whereas in cluster compounds strong
electron-phonon interaction will change the character of correlated hopping. 
These issues require a deeper analysis of our results.

We have reported here the metallic and insulating phases of Nb$_4$-cluster
compounds. The transport properties are dominated by electron localisation
and long-range Coulomb interactions among the carriers. The metallic phase
shows a minimum in resistivity. On the other hand, the insulating phase
exhibits hopping conduction with $\ln{\rho} \sim T^{-1/2}$ dependence,
suggesting the presence of a Coulomb gap at the Fermi level. These Nb$_4$
compounds show enhanced Pauli paramagnetism and their $\chi(T)$
behaviour, including a transition at about 30~K, is similar for the insulating as
well as metallic phases. Significantly, the absence of a corresponding anomaly 
in the transport behaviour suggests that in these compounds the transport properties
are unaffected by spin degrees of freedom.

{\bf Acknowledgements}: We thank Prof. Deepak Kumar for his comments and
suggestions. The magnetic measurements were performed during AN's
postdoctoral Visiting Fellowship at TIFR, Mumbai.

%%%%%%%%%%%%%%%%%%%%%%%FIGURES%%%%%%%%%%%%%%%%%%%%%%%%%%%%%%%

\newpage

%\newpage

%\newpage

%\newpage

%\newpage

\end{document}